\documentclass[aps,prx,twocolumn,letterpaper,groupedaddress,10pt]{revtex4-2}

\usepackage[utf8]{inputenc}
\usepackage{amssymb,amsthm,amsmath,amsfonts,mathptmx,mathcmd,mathrsfs}
\usepackage{graphicx}
\usepackage{dcolumn}
\usepackage{bm}
\usepackage{mathtools}
\usepackage[colorlinks=true,urlcolor=blue,citecolor=blue,linkcolor=blue]{hyperref}
\newcommand\Alpha{\mathrm{A}}
\newcommand\Beta{\mathrm{B}}
\DeclarePairedDelimiter{\bra}{\langle}{\rvert}
\DeclarePairedDelimiter{\ket}{\lvert}{\rangle}

\begin{document}

\title{Two Photon Tripartite Entanglement Transfer via Time-Multiplexed Quantum Walks}

\author{Jonas Lammers}
\author{Federico Pegoraro}
\author{Philip Held}
\author{Nidhin Prasannan}
\author{Benjamin Brecht}
\author{Christine Silberhorn}
\affiliation{
 Paderborn University, Integrated Quantum Optics, Institute for Photonic Quantum Systems (PhoQS), Warburger Stra\ss e 100, 33098 Paderborn, Germany
}

\date{\today}

\begin{abstract}
Photonic multidimensional quantum networks (MDQN), where individual subsystems are encoded using multiple degrees of freedom and photons, are an emerging platform for quantum algorithms because they offer high scalability. The distribution of non-classical and non-local correlations between the individual subsystems in an MDQN is of fundamental interest for many quantum protocols. Interestingly in an MDQN, the inseparability of two subsystems underlying entanglement can occur both between multiple distinct photons as well as between individual degrees of freedom associated with a single photon. In this work, we investigate the entanglement transfer enabled by the interplay of both entanglement between two distinct photons as well as inseparability between multiple degrees of freedom. For this purpose, we subject one photon of a polarization entangled two-photon pair to a discrete-time quantum walk introducing the position subsystem of the quantum walk as a third subsystem with qudit encoding. Here we study the resulting transfer of entanglement from the polarization degree of freedom, representing qubit encoding, towards the position degree of freedom, representing quidt encoding, via partial state tomography and correlation measurements.
\end{abstract}

\maketitle


Photonic quantum technologies are a rapidly developing field allowing for applications ranging from quantum communication \cite{6008516, Jacobs_2016, PhysRevLett.117.210501}, quantum computation \cite{Briegel2009, PRXQuantum.2.030325, doi:10.1126/science.aab3642} to quantum metrology \cite{Giovannetti2011, Pirandola2018, Zhang2021}. Many of these protocols build on entanglement in large photonic networks. In general, entanglement is understood in terms of the inseparability criterion where a system spanning two or more subsystems cannot be expressed via classical probability distributions over the individual subsystems. Typically these subsystems are considered to be identifiable single photons exhibiting a second degree of freedom to encode qubit/qudit information. This type of entanglement resulting in, for example, Bell, GHZ or Werner states, is typically referred to as particle entanglement and is considered a genuine quantum resource. While particle entanglement involving two qubits as the individual subsystems is a well understood and useful phenomena, entanglement involving multiple subsystems, higher dimensional encodings and or multiple degrees of freedom per single photon is vastly more complex. 


The platform of large scale multi-dimensional optical quantum networks (MDQN) incorporate both multiple photons and multiple degrees of freedom distributed between the individual parties involved. On the one hand, these MDQN feature particle entanglement as a genuine quantum resource. On the other hand, there exists a second form of inseparability in these MDQN. Mathematically similar in structure, one can define inseparability between multiple degrees of freedom on a single photon. The idea being that a system spanning multiple degrees of freedom is comprised of multiple subsystems each representing an individual degree of freedom. As the complete system becomes inseparable, in the sense that it cannot be decomposed into the tensor product of the individual subsystem, the complete system becomes entangled. This kind of single-photon entanglement, also referred to as modal \cite{PhysRevA.88.044301} or even classical \cite{Spreeuw1998ACA, PhysRevA.82.022115} entanglement, is considered as highly controversial \cite{doi:10.1126/science.aad7174}.

On its own, single-photon entanglement does not seem to be a genuine quantum resource, as the resulting probability distribution can be emulated using coherent laser light \cite{Spreeuw1998ACA, Aiello_2015}. This fact has sparked a series of classical experiments utilizing coherent light to demonstrate quantum like advantage \cite{Toeppel_2014, Berg-Johansen:15, PEREZGARCIA20151675, lpor.201500252}. However, once single-photon entanglement is combined with a second optical resource, quantum mechanical effects such as non-locality \cite{Ho_2014, PhysRevLett.92.180401, PhysRevLett.110.130401, PhysRevLett.66.252, PhysRevLett.75.2063} or quantum teleportation \cite{PhysRevLett.126.130502} can be observed. Recently, first theoretical studies by T. Giordani, et. al. \cite{Giordani_2021} have been conducted investigating the transfer between single-photon inseparability and genuine two-photon entanglement. They showed that using a MDQN it is possible to transfer and accumulate entanglement from a two-dimensional Bell-state towards a high-dimensional encoding using a MDQN, two-photons and four degrees of freedom. 

We therefore consider single-photon entanglement as a genuine resource which should not be discarded in MDQN networks. In this paper we experimentally demonstrate that it can be used to enrich the entanglement dynamics by transferring qubit-qubit entanglement towards qubit-qudit entanglement.
Specifically, we experimentally demonstrate entanglement transfer in a MDQN consisting of three subsystems spread across two parties (Alice \& Bob) each with their own spatially separated photon.
Alice has access to the polarization (subsystem $A_{\text{pol}}$) degree of freedom on her photon representing a qubit, while Bob has access to both the polarization (subsystem $B_{\text{pol/coin}}$) and high-dimensional time-bin (subsystem $B_{\text{time/pos}}$) degree of freedom on his photon representing a qubit and qudit respectively.
We generate both photons being initially entangled in their respective polarization degree of freedom and spatially separate them into Alice' and Bob's photon. Then Bob's photon undergoes a time-multiplexed quantum walk introducing the time-bin degree of freedom. Here we study the resulting transfer of entanglement between Alice and Bob's polarization towards Alice and Bob's position subsystem which never interact directly.
Note that this means that entanglement has been transferred from an entangled two qubit system towards and entangled qubit-qudit system.
In order to determine this transfer, we perform a partial reconstruction of the underlying density matrix to investigate generalized CHSH measures and perform correlation measurements between multiple bases ("remote conditioning", see \ref{sec:Concept}) in order to verify entanglement transfer.

\section{Framework}

In order to investigate entanglement transfer dynamics in a MDQN, we first introduce and contrast the notion of single-photon inseparability and two-photon entanglement by looking at correlations that can arise in a bipartite system in first and second quantization. We then introduce discrete-time quantum walks as our MDQN of choice followed by the experimental concept combining two-photon entanglement and single-photon inseparability as well as the utilized entanglement measures.

\subsection{Correlations}

Consider a bipartite system consisting of two subsystems, Alice ($A$) and Bob ($B$), spanning the combined Hilbert space $\mathcal{H} = \mathcal{H}_A \otimes \mathcal{H}_B$. We can now define a separable state $\hat{\sigma}$ as
\begin{equation}
    \hat{\sigma} = \int dP(\Psi_A, \Psi_B) \ket{\Psi_A, \Psi_B} \bra{\Psi_A, \Psi_B}
    \label{eq:Separable}
\end{equation}
where $P(\Psi_A, \Psi_B)$ represents a classical joint probability distribution \cite{PhysRevA.40.4277, PhysRevA.100.062129}. Using this notion of separable states, we can further identify the set of entangled states as those which cannot be expressed by 
Eq. (\ref{eq:Separable}), as they would require negativity from $P(\Psi_A, \Psi_B)$.

\begin{figure}[h]
    \includegraphics[width=0.4\textwidth]{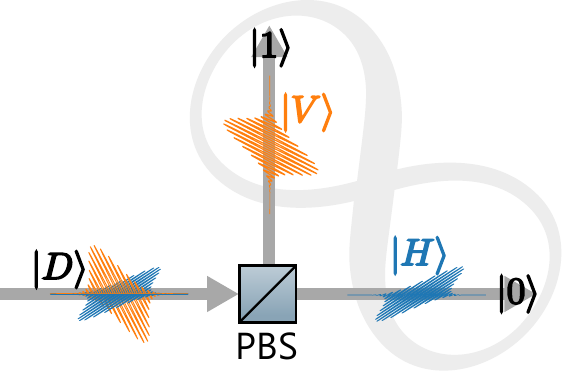}
    \caption{\label{fig:PBS}Example of single photon inseparability generation involving a spatial and polarization degree of freedom utilizing a polarization beam splitter (PBS).}
\end{figure}

An example of a bipartite system would be light impinging on a polarizing beam splitter (PBS) as displayed in Fig. \ref{fig:PBS}. Here we can identify Alice as the polarization $\left\{\ket{H}, \ket{V}\right\}$ and Bob as a two-dimensional position $\left\{\ket{0}, \ket{1}\right\}$ degree of freedom where $\ket{0}$ marks the transmitted and $\ket{1}$ the reflected path of the PBS. Now we can consider the situation where the light is initially diagonally $\ket{D}$ polarized.

In first quantization, we can describe the initial state as $\ket{\Phi} = \ket{D}_A\otimes\ket{0}_B$. A PBS can now separate the beam into the different spatial modes according to their respective polarization resulting in the state $\ket{\Psi} = 1/\sqrt{2} \left(\ket{H}_A\otimes\ket{0}_B + \ket{V}_A\otimes\ket{1}_B\right)$. While the initial state is clearly separable, there exists no classical joint probability distribution $P$ such that the resulting state $\ket{\Psi}$ can be expressed as a separable state. Effectively this shows how one can induce inseparability deterministically using linear optics in the first quantization picture.

Moving on to the second quantization, we can now describe the initial state for a single photon as $\ket{\Phi} = \hat{a}_{D,0}^\dagger\ket{\text{vac}}$ and the resulting state after the PBS as $\ket{\Psi} = 1/\sqrt{2} \left(\hat{a}_{H,0}^\dagger + \hat{a}_{V,1}^\dagger\right)\ket{\text{vac}}$ where $\hat{a}_{p,x}^\dagger$ is the creation operator acting on the polarization $p$ and position mode $x$. First of all, as this state consists of only a single photon, it is impossible to observe correlations between Alice and Bob. Therefore it is impossible to probe Bell non-locality with such a system which actually allows its simulation using classical coherent laser light \cite{Spreeuw1998ACA, Aiello_2015}. Secondly it has been shown \cite{PhysRevA.100.062129} that for any single-photon system exhibiting this kind of entanglement there exists a mode transformation capable of disentangling the system. However in contrast to coherent laser light, utilizing single-photon inseparability inside a MDQN where more than one photon is present, may allow for interesting dynamics as it can be combined with, for example, two-photon entanglement.

Finally, we examine two-photon entanglement. Specifically, we ignore the PBS as an optical nonlinearity is required in order to create two-photon entanglement and instead redefine Alice and Bob as having control over photons which can be identified via a spatial degree of freedom where we associate Alice with $\ket{0}$ and Bob with $\ket{1}$. Both photons possess a polarization degree of freedom which we utilize in order to create the Bell entangled state $\ket{\Psi}=1/\sqrt{2} \left(\hat{a}_{H,0}^\dagger \hat{a}_{V,1}^\dagger + \hat{a}_{V,0}^\dagger \hat{a}_{H,1}^\dagger \right)\ket{\text{vac}}$. This two-photon entanglement now allows for individual measurements of Alice' and Bob's subsystems and can therefore exhibit Bell non-locality.

In an MDQN, where multiple photons each with multiple degrees of freedom are present, the interplay between single photon inseparability and multi photon entanglement becomes of interest for the overall entanglement distribution in the network. The smallest interesting MDQN incorporating both effects is a two photon Bell-state where we introduce a high-dimensional degree of freedom, as a third party, on of the two photons. The resulting tripartite dynamics now allow for correlation measurements as well as performing entanglement transfer processes between the individual degrees of freedom. In this paper, we investigate the resulting entanglement transfer dynamics in exactly this scenario. For this purpose, we introduce the high-dimensional degree of freedom via a linear optical network realized by a discrete-time quantum walk.

\begin{figure*}
    \includegraphics[width=0.7\textwidth]{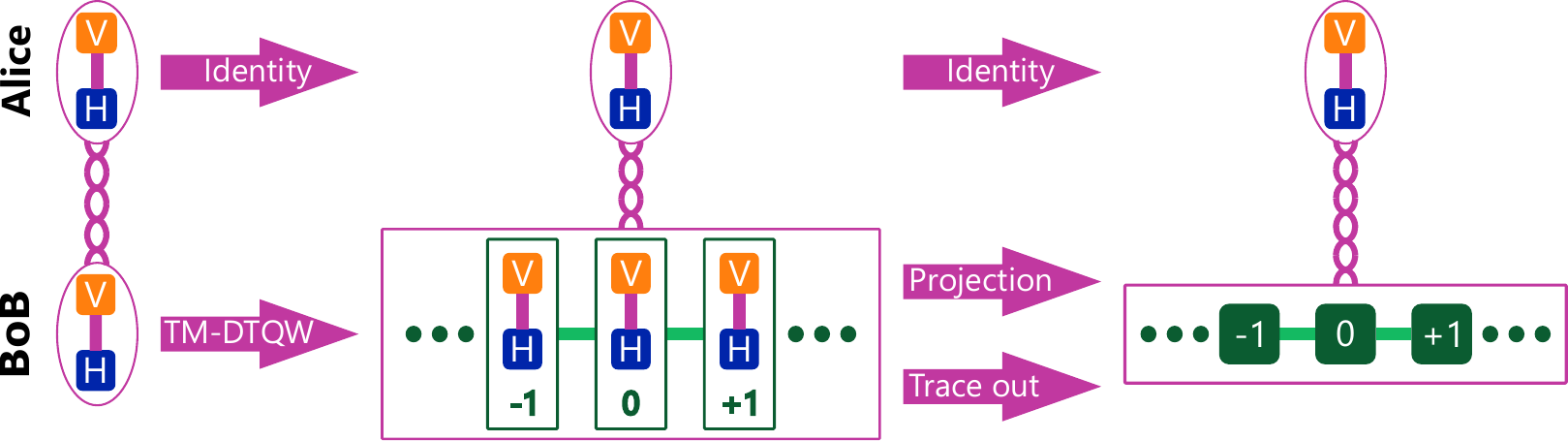}
    \caption{\label{fig:Concept}Conceptual idea of the experiment combing single photon inseparability (Bob's position \& Bob's coin) and two photon entanglement (Alice \& Bob) by creating two polarization entangled photons separated into Alice' and Bob's subsystem respectively. A third  position degree of freedom is introduced to Bob's photon via a TM-DTQW resulting in a tripartite state. The resulting entanglement between Bob's position and Alice' subsystem is investigated using projective measures and by tracing over individual subsystems.}
\end{figure*}

\subsection{\label{sec:DTQW} Discrete Time Quantum Walks}

Discrete time quantum walks (DTQWs) describe the evolution of one or many walkers on a graph structure. They are defined in a bipartite coin-position Hilbert space $\mathcal{H}_{c} \otimes \mathcal{H}_{x}$, where $\mathcal{H}_{x}$ denotes the position space implementing the graph vertices while $\mathcal{H}_{c}$ denotes the coin space which corresponds to the edges of the graph structure.
Therefore the state of a single photon walker on a one-dimensional lattice is defined as
\begin{equation} 
\ket{\Psi (t)} = \sum_{x, c} \Psi_{x,c}(t) \hat{a}^{\dagger}_{x,c}\ket{\text{vac}} = \sum_{x, c} \Psi_{x,c}(t) \ket{x,c} \in \mathcal{H}_x \otimes \mathcal{H}_c
\end{equation}
where $|\Psi_{x,c}(t)|^2$ indicates the probability of finding the walker at position $x \in \mathbb{Z}$ in coin state $c \in \{H,V\}$ at step $t \in \mathbb{N}_0$. For all $t$, $\sum_{x, c}|\Psi_{x,c}(t)|^2 = 1$ ensures normalized probability distributions.

The evolution of a DTQW occurs in discrete steps $t$ consisting of a coin operation $\hat{C}$ which in general performs an arbitrary rotation of the coin state at every position followed by a shift operation $\hat{S}$ updating the position state accordingly:
\begin{equation} 
\begin{aligned}
\hat{C} = 
    &\sum_{x \in \mathbb{Z}} \!\quad\! \sum_{p,q \in \{H,V\}} c_{p,q}(x) \ket{x}\!\bra{x} \otimes \ket{p}\!\bra{q}\\
\hat{S} = 
    &\sum_{x \in \mathbb{Z}} \left(
        \ket{x+1}\!\bra{x}\otimes \ket{H}\!\bra{H}
        + \ket{x-1}\!\bra{x}\otimes \ket{V}\!\bra{V} 
    \right)
\end{aligned}
\label{eq:stepoperator}
\end{equation}
The two together form one step of the unitary walk evolution resulting in $\hat{U}(t) = \left(\hat{S}\hat{C}\right)^t$. In order to illustrate the evolution of a DTQW we identify the two coin basis states with the vectors: $\ket{H}=(1,0)^T$ and $\ket{V}=(0,1)^T$. We further restrict ourselves to a Hadamard coin operation for every step and position, as implemented in the setup, resulting in the coin operation
\begin{eqnarray}
\hat{C}_H= \hat{\mathbb{I}}_x \otimes\frac{1}{\sqrt{2}} \begin{pmatrix}
1 &1\\
1 & -1
\end{pmatrix}.
\end{eqnarray}
As the coin and step operation are both unitary operations, the resulting evolution operator $\hat{U}(t)$ is a unitary operator for all $t$ and the final state $\ket{\Psi(t)}$ of a single walker stays pure if the initial state was pure. Theoretical investigations \cite{PhysRevLett.111.180503, chandrashekar2013} and experimental demonstrations \cite{Wang:18} have shown that a Hadamard DTQW causes the walkers position and coin degree of freedom to become inseparable.

\subsection{\label{sec:Concept} Concept}

In this work, we study entanglement transfer dynamics in a tripartite MDQN consisting of two spatially separated photons (Alice \& Bob) and three subsystems encoded in the polarization and time-bin degrees of freedom as displayed in Fig. \ref{fig:Concept}.
We associate one photon with Alice ($A$), who has access to its polarization degree of freedom (subsystem $A_{\text{pol}}$) representing a qubit.
The other photon we associate with Bob ($B$), wo has access to its polarization (subsystem $B_{\text{pol/coin}}$) as well as time-bin (subsystem $B_{\text{time/pos}}$) degree of freedom representing a qubit and qudit respectively.
We generate both Alice' and Bob's photon as part of a polarization encoded bell state and subject Bob's photon to a Hadamard DTQW, where we identify the DTQWs coin subsystem as $B_{\text{pol/coin}}$ and the DTQWs position subsystem as $B_{\text{time/pos}}$.
This results in the tripartite state scenario
\begin{equation}
    \begin{aligned}
        \mathcal{H} 
            &= \mathcal{H}_{A_{\text{pol}}} \otimes \mathcal{H}_{B_{\text{coin}}} \otimes \mathcal{H}_{B_{\text{pos}}} \\
            &= \text{span}\!\left\{
                \ket{c_A, c_B, x} : c_A,c_B \!\in\! \{H,V\} \!\land\! x \in\! \mathbb{Z}
            \right\} \\
            &= \text{span}\!\left\{
                \hat{a}^{\dagger}_{c_A}\hat{a}^{\dagger}_{c_B,x} \ket{\text{vac}} : c_A,c_B \!\in\! \{H,V\} \!\land\! x \in\! \mathbb{Z}
            \right\}
    \end{aligned}
\end{equation}
consisting of Alice' polarization $\mathcal{H}_A = \text{span}\{H, V\}$, Bob's coin $\mathcal{H}_{B_{\text{coin}}} = \text{span}\{H, V\}$  and Bob's position $\mathcal{H}_{B_{\text{pos}}} = \text{span}(\mathbb{Z})$ subsystem.
The resulting state of the combined system after $t$ DTQW steps can then be written as
\begin{equation}
    \label{eq:system_unitary}
    \ket{\Phi(t)} = \hat{\mathbb{I}}_A \otimes \hat{U}_B(t) \frac{\ket{H} \otimes \ket{V, 0} + \ket{V} \otimes \ket{H, 0}}{\sqrt{2}}.
\end{equation}
Using this process implemented on a MDQN, we investigate entanglement transfer between subsystems encoded as different degrees of freedom on a single photon ($B_{\text{coin}}$ \& $B_{\text{pos}}$) as well as parties with spatially separated photons (Alice \& Bob).
Specifically we study the transfer of the non-local two-photon entanglement between $A_{\text{pol}}$ and $B_{\text{coin}}$ towards $B_{\text{pos}}$.
This would result in high dimensional entanglement between $A_{\text{pol}}$ and $B_{\text{pos}}$ which were initially not entangled and never interacted directly.

In order to investigate entanglement transfer we first quantify the remaining entanglement between Alice' polarization and Bob's coin subsystem as the DTQW evolves.
For this purpose, we perform partial state tomography by reconstructing the systems density matrix $\hat{\rho}(x,t)$ individually at each step $t$ of the evolution and each position $x$ Bob's photon can occupy.
Using the reconstructed density matrices we calculate the generalized CHSH quantifier for entanglement
\begin{eqnarray}
    E(x,t) = \frac{\text{tr}\left[\hat{\rho}(x,t)\hat{L}\right] - 1}{4} 
        + \left|\frac{\text{tr}\left[\hat{\rho}(x,t)\hat{L}\right] - 1}{4}\right| \label{eq:CHSH} \\
    \text{where } \hat{L} = 
          L_x \hat{\sigma}_x \otimes \hat{\sigma}_x
        + L_y \hat{\sigma}_y \otimes \hat{\sigma}_y
        + L_z \hat{\sigma}_z \otimes \hat{\sigma}_z
\end{eqnarray}
where $\hat{L}$ is an optimized measure depending on the reconstructed density matrix $\hat{\rho}(x,t)$.
This quantifier scales between zero, indicating that the subsystems in question are not entangled, and one where both subsystems are maximally entangled.
Using this measure we study the average entanglement
\begin{equation}
    \label{eq:AverageEntanglement}
    E(t) = \sum_{x \in Z} P_t(x) E(x, t)
\end{equation} where $P_t(x)$ is the probability to detect Bob's photon at the position $x$ after $t$ steps. Under the assumption that the network performs a unitary evolution on Bob's photon, the entanglement between Alice' and Bob's parties should remain constant.
Therefore any decrease in the average entanglement means that not all entangled subsystems are accounted for in the quantifier.
This unaccounted party can be Bob's position subsystem, which means that the initial entanglement between Alice and Bob becomes distributed in the entire MDQN.
Therefore this quantifier can identify entanglement transfer processes in an MDQN.
However it cannot quantify with whom Bob's position subsystem becomes entangled.
Furthermore experimental imperfection can introduce further unaccounted parties.

In order to verify that entanglement is transferred inside the MDQN towards $A_{\text{coin}}$ and $B_{\text{pos}}$ we perform a second measure, we dub "remote conditioning".
The idea is to investigate the control Alice can exert over Bob's position subsystem using her remote polarization degree of freedom.
For this purpose we project Bob's coin subsystem on an arbitrary state $\ket{\Beta} = \text{cos}(\beta) \ket{H} + \text{sin}(\beta) \ket{V}$ on the equator of the Bloch sphere.
Alice performs an arbitrary polarization measurement on her photon $\ket{\Alpha} = \text{cos}(\alpha) \ket{H} + \text{sin}(\alpha) \ket{V}$ and Bob measures the corresponding position distribution $P_B(x, t)$ at every step $t$ of the MDQN.
Here we compare the control Alice can exert in the classically correlated scenario
\begin{equation}
    \hat{q}(t) = \hat{\mathbb{I}}_A \otimes \hat{U}_B(t) \frac{\ket{H, V, 0}\!\bra{H, V, 0} \!+\! \ket{V, H, 0}\!\bra{V, H, 0}}{2}
\end{equation}
with the control she can exert in the entangled scenario $\ket{\Phi(t)}$. Combined we measure Bob's position distribution
\begin{equation}
    \label{eq:RemoteConditioning}
    P_B(x, t) \!=\! \bra{A, B, x} \! \big[ (1 \!-\! \gamma) \hat{q}(t) \!+\! \gamma \ket{\Phi(t)}\!\bra{\Phi(t)} \big] \! \ket{A, B, x} 
\end{equation}
where $\gamma \in [0,1]$ switches between the classically correlated and quantum mechanically entangled scenarios and $\ket{A, B, x}$ projects Alice' and Bob's polarization state at position $x$ on $\ket{A}$ and $\ket{B}$ respectively.

\section{Experimental Setup}

\begin{figure*}
    \includegraphics[width=0.8\textwidth]{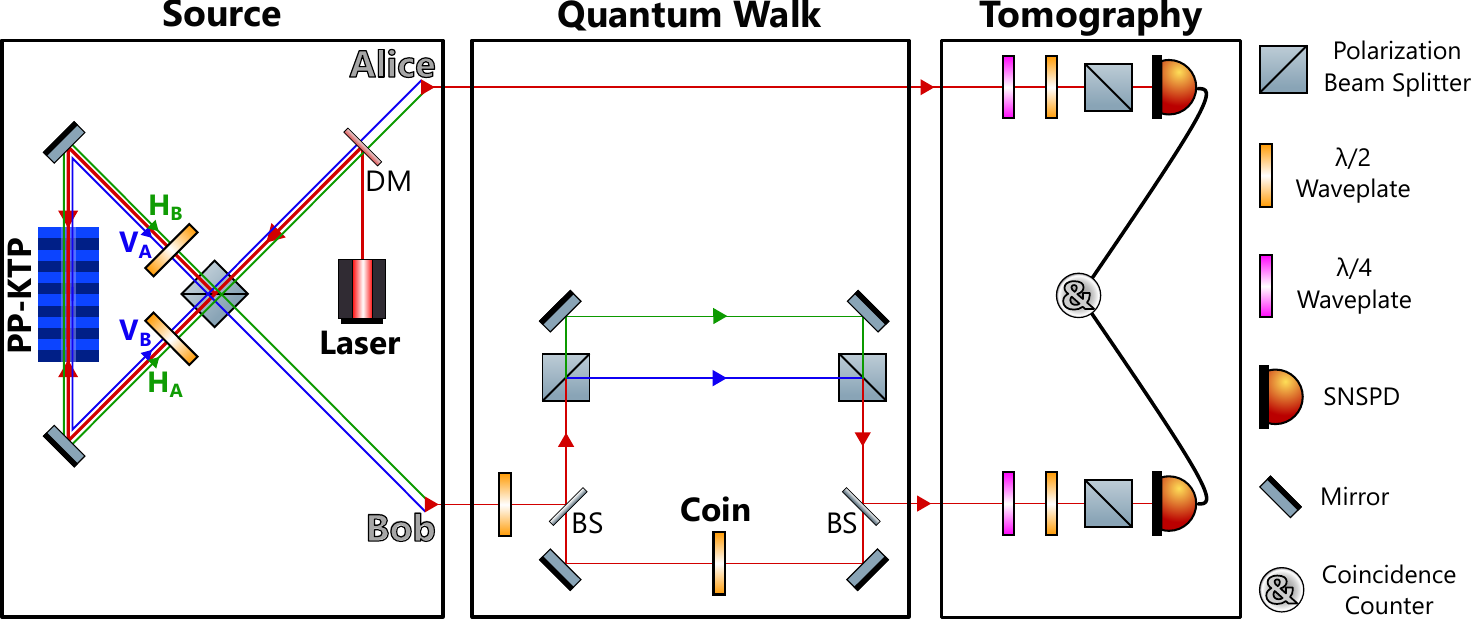}
    \caption{\label{fig:Setup} Sketch of the implemented setup consisting of a polarization entangled photon pair source in a Sagnac-Loop configuration (source), a time-multiplexed DTQW (Quantum Walk) implemented in free-space with probabilistic in- and out-coupling realized by beam splitters (BS) and polarization resolved two-photon detection (Tomography).}
\end{figure*}

The experimental setup is displayed in Fig. \ref{fig:Setup} consisting of the three sections: entanglement generation (source), time-multiplexed discrete time quantum walk (Quantum Walk) and detection (Tomography).

The source is a type-II spontaneous parametric down-conversion (SPDC) in periodically poled potassium titanyl (PP-KTP) waveguide, operated  in a Sagnac configuration \cite{MeyerScott2018HighperformanceSO}. Using a dichroic mirror, we pump the Sagnac loop with a diagonally polarized $770$ nm pulsed pump field with a $1.8$ nm bandwidth and $76$ MHz repetition rate. As the resulting clockwise and counterclockwise propagating down converted photon pairs interfere at the central polarization beam splitter (PBS), the which way information is erased. This results in the generation of two polarization entangled photons as the photon exiting one side of the PBS (Alice) has always the opposite polarization of the photon exiting the second port of the PBS (Bob). As characterized in \cite{MeyerScott2018HighperformanceSO} using HOM interference, we achieve a visibility of $(82 \pm 2)\%$ between two independently heralded photons. This limits the purity of the resulting entangled state to a maximum of $(84 \pm 1.6)\%$ and therefore the generated entanglement as quantified by Eq. (\ref{eq:CHSH}) to a maximum of $(73 \pm 3)\%$. We furthermore utilize a $1550$ nm laser as the classical reference for the remote conditioning measurement which we can insert in both Bob's and Alice' paths instead of the PDC-light after  the central PBS.

After the generation of the polarization entangled photon pair, we couple Bob's photon into our free-space DTQW implemented via the time-multiplexed architecture \cite{PhysRevLett.104.050502} using a partially reflective beam splitter. We implement the step operator $\hat{S}$ using an unbalanced Mach-Zehnder interferometer spanned by two PBSs which delays horizontally polarized light relative to vertically polarized light by $(125.4 \pm 0.1)$ ps. As both the horizontal and vertical parts are recombined on the same spatial mode after the unbalanced Mach-Zehnder interferometer, we implement the static coin operation $\hat{C}$ by placing a half-waveplate (HWP) into the beam path. From there the beam is directed back to the initial beam splitter completing a loop with a duration of $(3911.9 \pm 0.1)$ ps. This allows us to reuse the same unbalanced Mach-Zehnder and HWP to implement an arbitrary amount of quantum walk steps. Using a second partially reflective beam splitter between the unbalanced Mach-Zehnder and the HWP, we couple light with a probability of $(10.8 \pm 0.3)\%$ out of the loop at every position and every step of the DTQWs evolution and forward it towards the detection.

The detection unit of the setup consists of two superconducting nanowire single photon detector (SNSPD) channels with a nominal detection efficiency of $(82 \pm 3)\%$ and timing jitter of $(20 \pm 2)$ ps allowing us to resolve individual quantum walk positions. By restricting our measurements to correlations, where both detectors have clicked, we perform post-selection. Using a combination of quarter and half waveplates followed by a PBS, we can project both photons on arbitrary polarization states. This allows us to resolve all three involved degrees of freedom, further enabling us to perform two-photon polarization tomography for every position Bob's photon can occupy individually.

\bigskip

\section{Results}

\subsection{Entanglement Dynamics}

\begin{figure}[h]
    \includegraphics[width=0.45\textwidth]{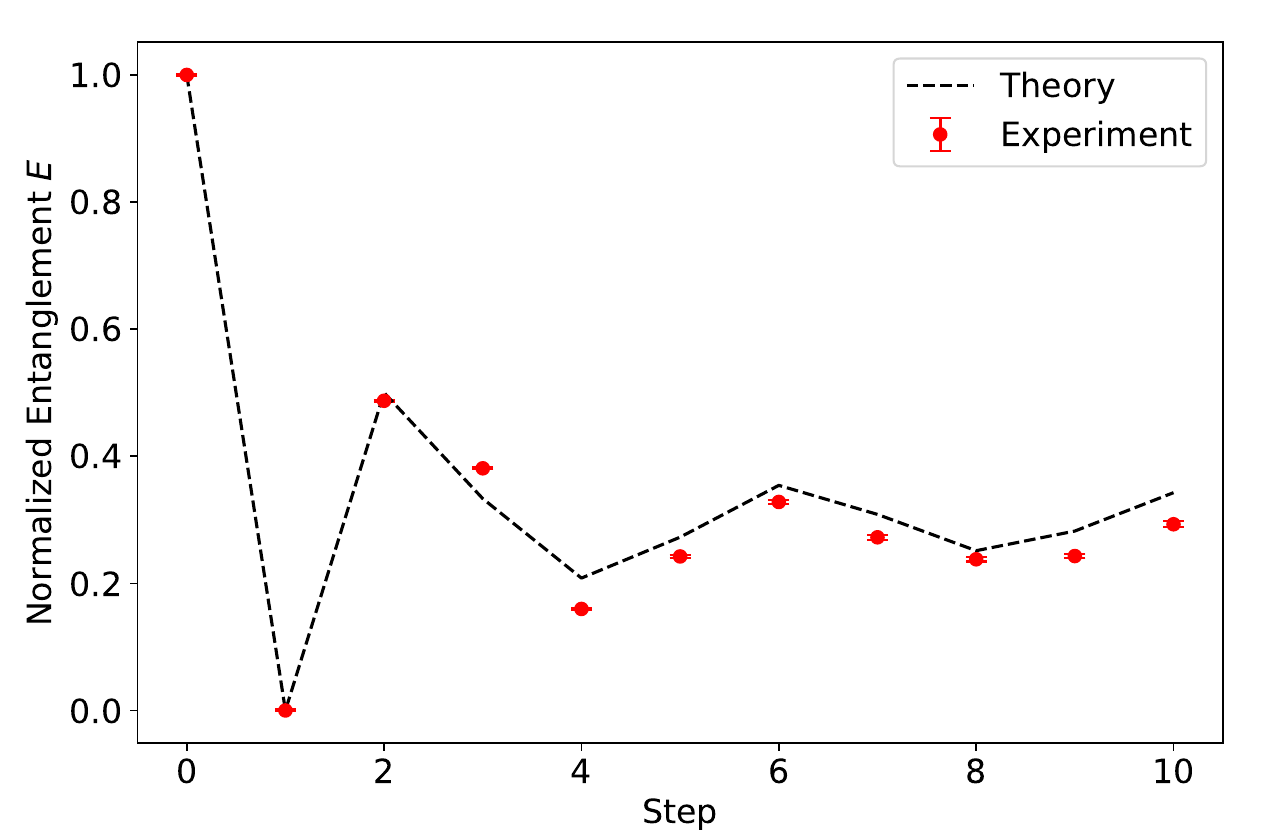}
    \caption{\label{fig:Entanglement} Comparison of the experimentally measured average entanglement $E(t)$ between Alice and Bob's polarization subsystem with theoretical predictions for the first 10 DTQW steps. The measured entanglement is normalized to the initial entanglement generated between Alice' and Bob's polarization encoded photons.}
\end{figure}

Resulting from the partial tomography, we reconstruct the density matrices $\hat{\rho}(x,t)$ describing the combined state of Alice' ($A_{\text{pol}}$) and Bob's ($B_{\text{coin}}$) polarization subsystem at every position $x$ and every step $t$ of the DTQW individually. For the first ten steps the reconstructed density matrices exhibit a purity $\text{tr}(\hat{\rho}(x,t)\hat{\rho}(x,t))$ of above $82\%$ and a similarity to theory of above $99\%$ with a few outliners where the probability for Bob's photon to occupy the respective position $x$ becomes exceedingly low. Using these reconstructed two photon polarization encoded density matrices, we now calculate the remaining entanglement $E(x,t)$ at each position $x$ and step $t$ between Alice and Bob's polarization subsystem. Next we calculate the average entanglement at each step $E(t)$ as discussed for Eq. (\ref{eq:AverageEntanglement}). Initially, at step zero, our system exhibits an entanglement of $(64.88 \pm 0.13)\%$ between Alice' and Bob's polarization encoded photons. We then normalize our measured entanglement in respect to this initial entanglement as we are interested in the transfer of this initial entanglement throughout the MDQN. The resulting normalized average entanglement is displayed in Fig. \ref{fig:Entanglement}.

After performing a single DTQW step, we can see that the entanglement between Alice photon and Bob's polarization subsystem has completely vanished, reaching a value of $(0.00 \text{}^{+0.15}_{-0.00})\%$ as theoretically predicted. As we apply only local operators, this means that the entanglement has been completely transferred towards a third party. This third party can now be either Bob's position subsystem or some unknown party in our experimental system. In step two, we can see that the normalized entanglement between Alice photon and Bob's polarization subsystem recovers reaching $(48.73 \pm 0.19)\%$ at step one. As this recovery occurs due to a local operation on Bob's photon, it can only be that entanglement has been transferred from some local third party back to Bob's polarization degree of freedom. This confirms that the entanglement has been transferred towards this local third party coherently. From there we see that the entanglement between Alice' and Bob's polarization subsystems follows theoretical predictions quite closely with an oscillating behavior around $(29 \pm 9)\%$. This means that as we perform more and more steps the dimensionality of Bob's subsystems increases while the amount of transferred entanglement stays approximately constant. As our system is highly scalable due to the time-multiplexing architecture, this effectively means that we expect to transfer around $70\%$ of the initial entanglement between Alice and Bob's polarization subsystem towards the rest of the MDQN where Bob's position subsystem can become arbitrarily large. As a next step, we investigate if this means that Alice becomes entangled with Bob's position subsystem at a distance without both systems having interacted directly.

\subsection{Remote Conditioning \label{sec:results:RC}}

\begin{figure}[h]
    \includegraphics[width=0.5\textwidth]{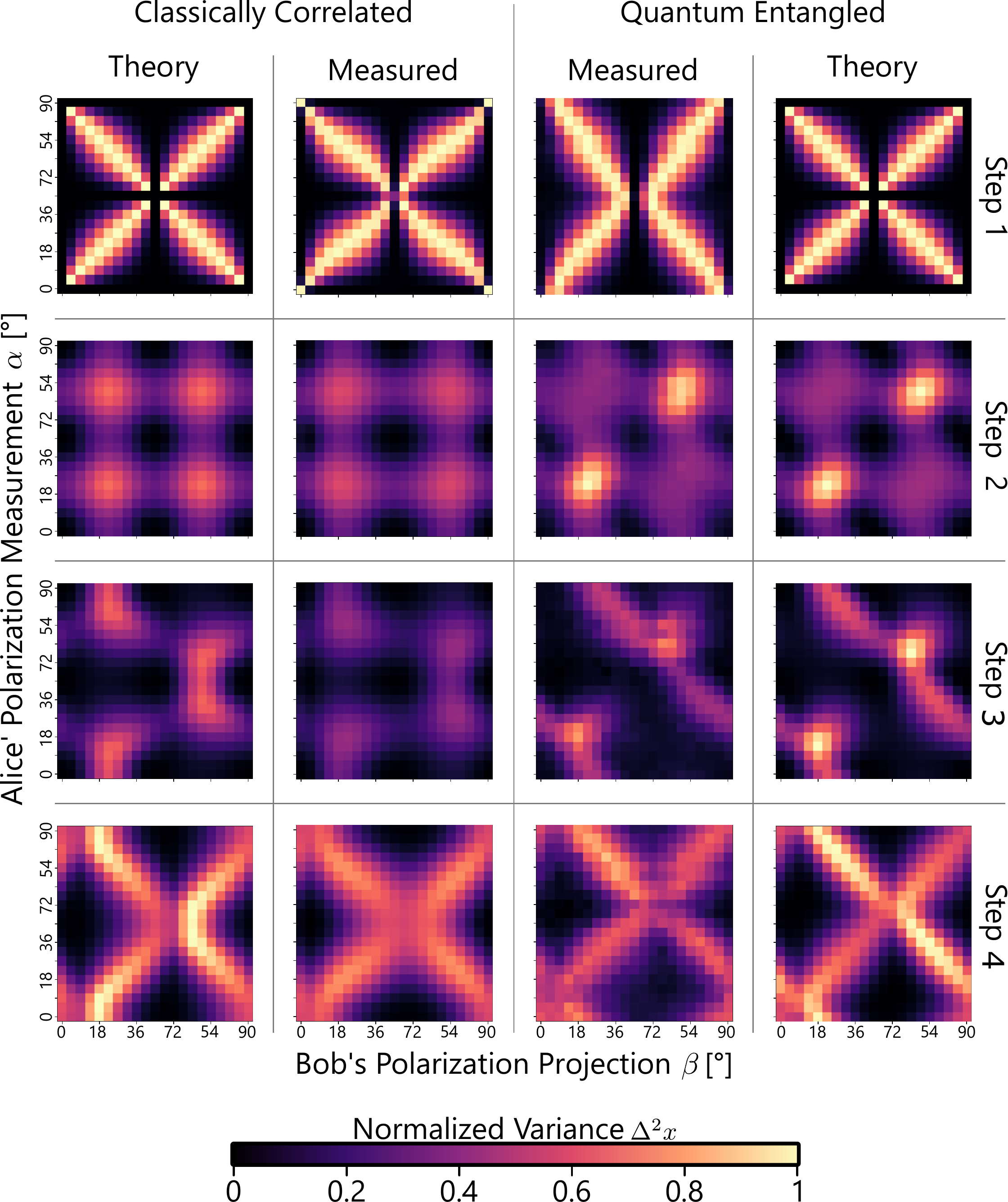}
    \caption{\label{fig:RemoteConditioning} Simulated and measured variances $\Delta^2x$ of the average position of Bob's position distribution after $n$ DTQW steps for both classically correlated and entangled multi particle states depending on the projection angle of both Alice' and Bob's polarization DoF. The displayed variances are normalized between the minimum and maximum measured or predicted variances for each step individually.}
\end{figure}

An experimentally feasible way to study the transfer of entanglement in our MDQN towards Alice' polarization ($A_{\text{pol}}$) and Bob's position ($B_{\text{pos}}$) subsystem is to probe the non-local control Alice can exert over $B_{\text{pos}}$. As discussed for Eq. (\ref{eq:RemoteConditioning}), the idea is to project Bob's polarization subsystem on an arbitrary polarization along the equator of the Bloch sphere. This reduces our MDQN to a two party system consisting only of $A_{\text{pol}}$ and $B_{\text{pos}}$. In this bipartite system we investigate what effects Alice' polarization measurement on her remote photon has on Bob's position distribution $P_B(x,t)$. In order to reduce the complexity of this investigation, we reduce the information of Bob's position subsystem into its variance $\Delta^2x$, a well established feature of DTQWs. Finally, we compare the control Alice can exert over the resulting variance $\Delta^2x$ of $B_{\text{pos}}$ in a classically correlated ($\gamma = 1$) and entangled ($\gamma = 0$) scenario. This comparison is displayed in Fig. \ref{fig:RemoteConditioning} including theoretical predictions for $\gamma = 0.2$ matching the expected mixedness as measured from the HOM dip visibility and $\gamma = 1$ for a state exhibiting only classical correlations.

Here we observe that after a single shift operation ($\hat{S}$) is performed at step one, Alice can exert the same control over Bob's position distribution, both in the considered classically correlated and entangled scenario. This observation matches our theoretical exception, due to our choice of the considered classically correlated state. Specifically, we note that at step one, where the DTQW applies a single shift operation ($\hat{S}$), Bob's horizontal polarization is shifted to a later position, while his vertical-polarization is shifted to an earlier position. This effectively means that any coherence term between Bob's horizontal and vertical polarization can not effect the project on each individual position of $B_{\text{pos}}$. Therefore, we know that our classically correlated state matches the entangled scenario up to the coherence term at step one.

Once another coin and shift operation has been performed at step two, the variance $\Delta^2x$ of Bob's position subsystem exhibits clear structural differences between a classically correlated and entangled scenario depending on which measurement Alice has performed on her photon. Especially striking is that at step two and three the maximum variance measured in the entangled scenario exceeds the variance in the classically correlated scenario by a significant margin. Specifically, we measure a variance of $(3.24 \pm 0.03)$ in the entangled scenario at step two, while the classically correlated one only reaches $(1.93 \pm 0.01)$. Note that in theory, one can construct classically correlated states reaching a maximum variance of $4$ as shown in the appendix. However, as the necessary classically correlated state would produce a different pattern at step one, we can rule it out as a possibility and therefore state that there exist no classically correlated state which matched both step one and step two. This argument becomes only stronger as we consider more and more quantum walk steps. Consequently we have verified that the non-local control Alice can exert over Bob's position subsystem depends on the entanglement present in the system which therefore demands that entanglement has been transferred towards the bipartite subsystem of $A_{\text{pol}}$ and $B_{\text{pos}}$ which never interacted directly.

\section{Discussion}

In this work we investigated the entanglement transfer inside a MDQN. Initially two parties (Alice \& Bob) encoded on two distinguishable photons exhibiting a polarization degree of freedom are entangled. Using a discrete-time quantum walk, we introduce a high dimensional third subsystem to our MDQN encoded as time-bins on Bob's photon. Here we studied, how the initial entanglement transfers inside the MDQN and which parties become entangled. Specifically, we monitor the remaining entanglement between Alice' and Bob's polarization encoded qubits as the DTQW evolves. We see that approximately $70\%$ of the initial entanglement is transferred towards the remaining MDQN. Using a remote conditioning method, we have also shown that some of that entanglement is transferred towards the bipartite subsystem spanning Alice' polarization encoded qubit and Bob's position encoded qudit. These results demonstrate that MDQN, in which parties encoded on multiple photons and multiple degrees of freedom, can exhibit complex entanglement dynamics, where even multiple parties encoded on a single photon can contribute to non-local and non-classical correlations. This paves a way towards the utilization of composite quantum network consisting of multiple photons as well as multiple degrees of freedom.

An interesting next steps would be to combine two-photon and single-photon dynamics in a quantum network comprised of multiple degrees of freedoms. For example by combining two photons from two independent polarizing entangled photon pairs in a quantum walk, one could combine both two-photon dynamics and single-photon dynamics. This would further allow for the exploitation of the resulting entanglement transfer to achieve a high degree of control over the two-photon dynamics in the quantum walk via the two remaining remote photons.

\section{Acknowledgments}

This work has received financial support by the European Commission through the Horizon Europe project EPIQUE (Grant No. 101135288). The authors thank Jan Sperling and Laura Ares for their continued involvement in developing the underlying idea for the experiment as well as many fruitful discussions.

\section{Author Contribution}

J.L. performed the experiment and analyzed the data. F.P. and P.H. provided experimental support and assisted in developing the theoretical framework. N.P. operated the source. J.L. wrote the manuscript. C.S. and B.B. supervised the work. C.S. initiated the work. All authors discussed the results and commented on the manuscript.

\section{Appendix}

\begin{figure*}
    \includegraphics[width=0.65\textwidth]{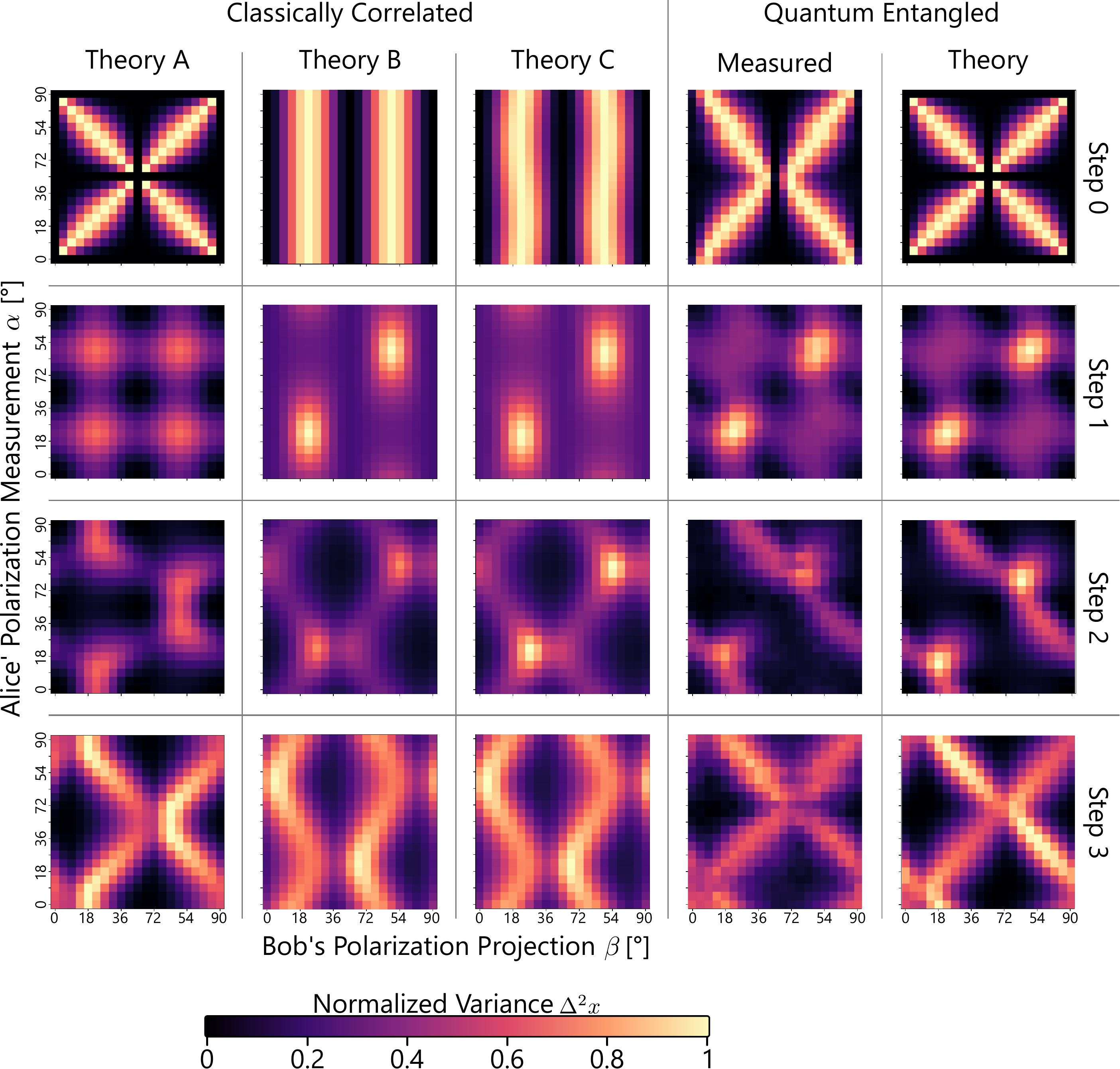}
    \caption{\label{fig:RemoteConditioningAppendix} Simulated and measured variances $\Delta^2x$ of the average position of Bob's position distribution after $n$ DTQW steps for both classically correlated and entangled multi particle states depending on the projection angle of both Alice' and Bob's polarization DoF. The displayed variances are normalized between the minimum and maximum measured or predicted variances for each step individually.}
\end{figure*}

In the following, we will expend on the effects of the choice of classically correlated state for the remote conditioning. In section \ref{sec:results:RC}, we showed that classical correlated states are insufficient in order to reproduce the remote conditioning dynamics under investigation. Specifically, we showed that the considered classical correlated state diverges from our measurement after step one of the DTQW, but matches it for step zero. Here we now show some further examples of classically correlated states which match the recorded dynamics at some step two and three for some angles, specifically showing that we can reproduce the same extreme points using classically correlated states. For this purpose, we choose the linear (Theory A), diagonal (Theory B) and one at $24^{\circ}$ (Theory C) polarization basis to represent our initial state and then remove the coherence terms between Alice and Bob, resulting in a statistical mixture. The resulting remote conditioning dynamics are displayed in Fig. \ref{fig:RemoteConditioningAppendix}, where we can see that we can reach the same extreme points. However, these states still fail to capture the complete dynamics, reinforcing our argument that entanglement must be transferred.

\bigskip

\bibliography{biblio}

\end{document}